\documentclass[manuscript]{aastex}

\usepackage{emulateapj5}

\newcommand\spie{{Proc.\ SPIE}}

\slugcomment{Accepted by AJ for February 2003}

\shorttitle{2MASS 1503+2525}
\shortauthors{Burgasser et al.}

\begin{document}

\title{The 2MASS Wide-Field T Dwarf Search. I. Discovery of a Bright T Dwarf
Within 10 pc of the Sun}

\author{
Adam J.\ Burgasser\altaffilmark{1,2}, J.\ Davy
Kirkpatrick\altaffilmark{3}, Michael W.\ McElwain\altaffilmark{1},
Roc M.\ Cutri\altaffilmark{3}, Albert J.\
Burgasser\altaffilmark{4}, \& Michael F.\
Skrutskie\altaffilmark{5}}

\altaffiltext{1}{UCLA Division of Astronomy \& Astrophysics, 8965
Math Science Bldg., 405 Hilgard Ave., Los Angeles, CA, 90095-1562;
adam@astro.ucla.edu, mcelwain@astro.ucla.edu}
\altaffiltext{2}{Hubble Fellow}
\altaffiltext{3}{Infrared Processing and Analysis Center, M/S
100-22, California Institute of Technology, Pasadena, CA 91125;
davy@ipac.caltech.edu, roc@ipac.caltech.edu}
\altaffiltext{4}{John J.\ McCarthy Observatory, P.O.\ Box 1144,
New Milford, CT 06776; aburgasser@mccarthyobservatory.org}
\altaffiltext{5}{University of Virginia,
Department of Astronomy, P.O.\ Box 3818, Charlottesville,
VA 22903; skrutskie@virginia.edu}

\begin{abstract}
We present the discovery of a bright (J = 13.94$\pm$0.03) T dwarf,
2MASS 1503+2525, identified in a new, wide-field search for T
dwarfs using the recently completed Two Micron All Sky Survey
(2MASS). The 1--2.5 $\micron$ spectrum of this object exhibits the
strong H$_2$O and CH$_4$ bands characteristic of mid- and
late-type T dwarfs, and we derive a spectral type of T5.5 using
both the Burgasser et al.\ and the Geballe et al.\ classification
schemes. Based on its spectral type and the absolute magnitudes of
known T dwarfs, we estimate the distance of this object as 8$\pm$3
pc if it is single, likely within 10 pc of the Sun. Our new 2MASS
search, which covers 74\% of the sky and greatly expands on
earlier color constraints, should identify 15--25 new T dwarfs
with J $\leq$ 16. Combined with the 20 known members of this class
that already fall within our search criteria, our new sample will
provide improved statistics for such key quantities as the binary
fraction and the field substellar mass function.  Furthermore,
multiple detections from overlapping 2MASS scans provide multiple
epoch astrometry and photometry, and we present proper motions for
five T dwarfs in our sample.
\end{abstract}


\section{Introduction}

T dwarfs are low-temperature (T$_{eff}$ $\lesssim$ 1300--1500 K)
brown dwarfs exhibiting distinct absorption bands of CH$_4$ in the
near-infrared H- and K-bands \citep{kir99,me02,geb02}. They are
distinguished from warmer L-type \citep{kir99,mrt99} and M-type
dwarfs in the near-infrared by the presence of these molecular
features, in addition to significant H$_2$O and collision-induced
(CIA) H$_2$ absorption \citep{sau94}. Since the discovery of the
prototype of this class, Gliese 229B \citep{nak95,opp95}, over 30
T dwarfs have been identified, most in the wide-field Two Micron
All-Sky Survey \citep[hereafter 2MASS]{skr97} and Sloan Digital
Sky Survey \citep[hereafter SDSS]{yor00}. Classification schemes
for these objects based on near-infrared spectral morphology have
been defined by \citet{me02} and \citet{geb02}.

T dwarfs identified in wide-field surveys are all likely within 25
pc of the Sun, the limiting distance for the NASA/NSF NStars
Project. Indeed, brown dwarfs in general may constitute a
significant population in the Solar Neighborhood, roughly equal in
number to the stellar population \citep{rei99,cha02}.  They have
generally eluded detection, however, because of their intrinsic
faintness. As the Solar Neighborhood is the basis for many studies
of stellar atmospheres, star formation and evolution, initial mass
function, and Galactic structure, characterizing the local
substellar population is of some importance.  This is especially
true at closer distances, such as the 10 pc RECONS sample
\citep{hen97}, where the stellar census is far more complete, but
few L- or T-type stars and brown dwarfs are currently known.

In this article, we present the discovery of a bright T dwarf,
2MASS J15031961+2525196\footnote{Throughout this paper, we cite
designations as they appear in the final 2MASS Point Source
Catalog (PSC), given as ``2MASS Jhhmmss[.]ss$\pm$ddmmss[.]s''. The
suffix conforms to IAU nomenclature convention and is the
sexagesimal Right Ascension and declination at J2000 equinox.}
(hereafter 2MASS 1503+2525), identified in a new search of the
all-sky 2MASS Working Point Source Database (WPSD).  The
spectrophotometric distance of this object places it within 10 pc
of the Sun. In $\S$2 we describe our new wide-field T dwarf search
and compare it to previous search efforts.  In $\S$3 we discuss
the identification of 2MASS 1503+2525 in our search and its
subsequent confirmation as a T dwarf using the Lick 3m Gemini
Infrared Camera. In $\S$4 we derive the spectral type and
spectrophotometric distance of 2MASS 1503+2525. We discuss our
results in $\S$5.

\section{The 2MASS Wide-Field T Dwarf Search}

Our current program is a continuation of a previous 2MASS search
\citep{me02} conducted when this survey was roughly 50\% complete.
2MASS is ideal for identifying T dwarfs, as the JHK$_s$ filters it
has employed sample the peak of the spectral energy distribution
of cool brown dwarfs, enabling us to probe relatively deeply and
over a large area of sky.  A total of 17 T dwarfs were detected in
our earlier search in three separate samples: a primary sample
covering 16,620 deg$^2$, requiring J $<$ 16, J--H $<$ 0.3, and
H--K$_s$ $<$ 0.3 for candidates selected from the 2MASS WPSD; and
two auxiliary samples selected from 2MASS Incremental Release Data
with smaller search areas but somewhat different color criteria.
However, the overall stringent color cuts used in this search
prevented our detecting early- and mid-type T dwarfs
\citep{leg00}, while our small sample resulted in large
statistical uncertainties for population statistics such as the T
dwarf space density and substellar mass function (Burgasser 2001;
Burgasser et al.; in preparation).

Our current search effort, using the 2MASS WPSD, attempts to
address these shortcomings in two ways. First, we have expanded
our color criteria to J--H $\leq$ 0.3 {\em or} H--K$_s$ $\leq$ 0,
allowing detection of objects with redder near-infrared colors
(and hence earlier spectral type), or faint objects with larger
photometric uncertainties that were not identified in our first
primary sample.  These color constraints are required to exclude
an overwhelming number of background stars, as discussed below,
and effectively limit our search to spectral types roughly T4 and
later \citep{me02,leg02}. We continue to impose signal-to-noise
(S/N), photometric completeness, and reliability constraints by
selecting only those candidates with J $\leq$ 16 (S/N $\sim$ 10
completeness limit) and both J- and H-band detections. Second,
with the recent completion of 2MASS survey operations, we are now
able to search the entire sky for T dwarfs, excluding only those
regions with very high source densities and hence likelihood of
source confusion: the Galactic Plane (${\mid}b{\mid} < 15\degr$),
the Large and Small Magellanic Clouds, and various dense or
obscured stellar regions such as 47 Tuc and the Orion Nebular
Cluster. We have also excluded the Equatorial poles
(${\mid}{\delta}{\mid}
> 88\degr$) because of restrictions on observational followup. In
total, our current search encompasses 30,400 deg$^2$, or 74\% of
the sky, increasing our previous sample area by over 80\%. We
expect to find roughly 15 new T dwarfs in this sample based on the
increased areal coverage alone, and up to 10 additional red and/or
faint T dwarfs based on the number of objects found in the
\citet{me02} auxiliary samples.

In Figure 1 we plot the J--H and H--K$_s$ Hess diagram
\citep{hes24} of our 264,068 2MASS-selected candidates, selected
from over 1.3 billion sources in the 2MASS WPSD, along with the
2MASS colors of T dwarfs detected by 2MASS and SDSS (Table 7 in
Burgasser et al., 2002; and Table 4 in Geballe et al., 2002; all
photometry from 2MASS). The
density of candidates is highest near our color constraints,
particularly at the early-type end of the \citet{bes88} dwarf
track and near the turnoff between the dwarf (late-K to early-M
type) and giant tracks. The vast majority of these candidates are
likely faint background stars, which are densely centered at and
around the \citet{bes88} tracks.  While our 2MASS-selected
candidates are required to have no optical counterparts within
5$\arcsec$ in the USNO A2.0 catalog \citep{mon98}, based on the
observed extreme red colors of known T dwarfs (R--J $\gtrsim$ 9;
Golimowski et al.\ 1998), a substantial number of faint sources,
proper motion stars, closely-separated visual binaries, and
optically red sources (the USNO A2.0 catalog requires detections
in both R and B bands) remain.  Hence, we visually examine
multiple-epoch Digital Sky Survey (DSS) images of all candidate
fields to exclude these contaminants.  Of the roughly 70,000
candidates so far examined, 99.5\% have been eliminated in this
manner.  The remaining steps of our search process, as discussed
in \citet{me02}, include follow-up near-infrared imaging for the
elimination of minor planet contaminants, optical imaging to
exclude faint background stars, and near-infrared spectroscopy for
final identification. These follow-up observations are currently
underway.

Our color and magnitude constraints encompass 20 T dwarfs
previously identified by 2MASS and SDSS, listed in Table
1\footnote{Gliese 229B is not listed in Table 1, for
at a separation of 7$\farcs$8 \citet{nak95} it is obscured
by its brighter primary in the 2MASS imaging data.
Distant T dwarfs, such as NTTDF
1205$-$0744 \citep{cub99}, IFA 0231$-$Z1 \citep{liu02},
and SOri 0538$-$0236 \citep{zap02} are too
faint to satisfy our magnitude constraints.}. This sample is
already larger than that studied by \citet{me02}. Furthermore,
seven of these sources have multiple detections in the 2MASS WPSD
due to overlapping or repeated scans, as listed in Table 2. These
observations, separated by as much as 2 years, enable us to derive
mean proper motions for six of the T dwarfs\footnote{The proper
motion for Gliese 570D has been previously reported in
\citet{me00a}, confirming that object's association with Gliese
570ABC.}, assuming positional uncertainties of
0$\farcs$3\footnote{Characterization of the astrometric and
photometric properties of 2MASS point sources can be found in
Cutri et al.\ (2002);
\url{http://www.ipac.caltech.edu/2mass/releases/second/doc/explsup.html}.}.
Note that these values have not been corrected for Solar motion.
Four of the objects, 2MASS 0937+2937, SDSS 1346$-$0031, Gliese
570D, and 2MASS 2339$-$1352, have motions of at least 0$\farcs$5
yr$^{-1}$.  The substantial motions of these objects are not
unexpected, given that T dwarfs detected by 2MASS are all likely
within 20 pc of the Sun.

A comparison of the multiple-epoch photometry for the objects
listed in Table 2 gives some hint of photometric variability.
Reasonably significant (maximum $\Delta$M $\gtrsim$ 2$\sigma$)
variations are seen at J-band for SDSS 0926+5847 (2.8$\sigma$),
2MASS 1237+6526 (2.1$\sigma$), Gliese 570D (2.8$\sigma$), and
2MASS 2339+1352 (3.6$\sigma$); and at H-band for SDSS 1346$-$0031
(2.3$\sigma$).  It is possible that these photometric variations
are intrinsic, as has now been observed for a number of brown
dwarfs \citep[Enoch, Brown, \& Burgasser, in
preparation]{bai99,bai01b,cla02,gel02}.  \citet{me02b} have hypothesized
that these variations arise from the motions and evolution of
patchy condensate clouds in the upper atmospheres of early- and
mid-type T dwarfs; indeed, the variations appear to be greatest at
J-band, as predicted, and possibly more pronounced in the T4.5
SDSS 0924+5847 and T5.5 2MASS 2339+1352 (note the equally strong
variation in the T8 Gliese 570D, however.)  Such low-level
photometric variability as measured by the 2MASS J-band filter
must be considered carefully, however, as the filter bandpass
extends well into the 1.4 $\micron$ telluric H$_2$O band.  This
results in substantial variations in the J-band zeropoint (up to
0.1 mag) from night-to-night and within a single night (Cutri et
al.\ 2002).  The source of these fluctuations, likely changes in
observing conditions (terrestrial weather and water vapor), can
potentially affect the photometric calibration of late-L and T
dwarfs, as their spectral energy distributions show far more
structure in the J-band than earlier-type calibration stars. Such
second-order extinction effects could therefore be responsible for
the apparent variability seen, and a more directed program is
required to validate these observations. Regardless, our complete
sample will provide time-resolved astrometric and potentially
photometric data for a few individual T dwarfs, in addition to
improved population statistics.

\section{Identification of 2MASS 1503+2525}

\subsection{Initial Selection}

2MASS 1503+2525 is one of the first objects examined in our new
wide-field search campaign.  Astrometry and photometry for this
object are listed in Table 1, while Figure 2 shows the 2MASS
1503+2525 field as imaged by POSS-I \citep{wil52}, POSS-II
\citep{rei91}, and 2MASS.  This object is one of our brightest
candidates, just fainter than the brightest known T dwarf 2MASS
0559$-$1404 \citep{me00c}.  The absence of an optical source at
the position of 2MASS 1503+2525 implies R$-$J $\gtrsim$ 6
\citep{rei91}, while its near-infrared colors (J--H =
0.08$\pm$0.05, H--K$_s$ = $-$0.11$\pm$0.08) are similar to the
T6.5 Gliese 229B \citep{leg99}. The optical data also rule out the
possibility that this candidate is an earlier-type high-velocity
proper motion star, as no such counterpart is seen at either
epoch. Note that the bright star 18$\arcsec$ North and
103$\arcsec$ East of 2MASS 1503+2525 is the M5 variable star EX
Boo (a.k.a.\ IRAS 15012+2537, BD+25 2864, HIP 73662), classified
as a normal giant/supergiant by \citet{gug97}. A lower limit on
this object's Hipparcos distance, d $>$ 680 pc \citep{pry97},
eliminates the possibility that it is associated with the T dwarf.

\subsection{Lick Gemini Infrared Camera Observations}

We imaged 2MASS 1503+2525 in the near-infrared on 27 June 2002
(UT) using the Gemini Twin-Arrays Infrared Camera \citep{mcl93},
mounted on the Lick Observatory Shane 3m Telescope. Light cirrus
was present and seeing was 1$\farcs$2. Short integration (20 sec),
J- and K-band images unambiguously confirmed the presence of 2MASS
1503+2525 at its 2MASS position, and no secondary component was
seen at the seeing spatial resolution of the data.

We subsequently obtained low-resolution
(${\lambda}/{\Delta}{\lambda}$ $\sim$ 500) JHK spectra using the
Gemini Camera's grisms on two separate nights, 28 June 2002 and 19
August 2002 (UT).  Conditions for the first night (J- and K-band
spectra) were affected by heavy cirrus and 2$\arcsec$ seeing,
while smoke and light cirrus (1$\farcs$5 seeing) were present on
the second night (H-band spectrum).  For both sets of
observations, 2MASS 1503+2525 was first acquired in imaging mode
and placed into a 1$\farcs$4 slit. Sets of nodded (30$\arcsec$)
pairs were obtained with individual integration times of 180--300
sec, for total integrations of 1800 sec at J- and K-bands, and
2000 sec at H-band. Nod pair sets of the F8 V HD 133460 were
obtained for calibration immediately after the target observations
and at similar airmass.

Spectral images were pairwise subtracted; divided by a normalized,
median-combined set of flat-field quartz lamp observations; and
corrected for bad pixels by linear interpolation. Spatial
curvature was measured using the calibrator star dispersion lines,
and spectral curvature was measured from Argon lamp spectra
obtained during each of the observing runs. These lamp spectra
were also used for wavelength calibration, using the
identifications of \citet{stv68}. Target and calibrator spectra at
each nod position were then individually extracted and
interpolated onto a common wavelength scale. Relative flux
calibrations and telluric corrections were made by dividing target
spectra by calibrator spectra at each nod position, interpolating
over Paschen and Brackett Hydrogen lines. The resulting individual
spectra for each band were multiplied by a 6095 K blackbody
\citep[corresponding to spectral type F8 V]{tok00} and averaged.
2MASS magnitudes were used to flux calibrate the separate bands by
the prescription of \citet{me02}.

Figure 3 plots the reduced Gemini Camera grism spectrum of 2MASS
1505+2525. Signal-to-noise ratios are roughly 17 at J, 13 at H,
and 9 at K. CH$_4$ bands at 1.15, 1.4, and 2.2 $\micron$, and
H$_2$O bands at 1.1, 1.3, and 1.9 $\micron$ are clearly evident,
as are the 1.25 $\micron$ K I doublet lines.  The 1.17 $\micron$ K
I doublet lines are obscured by noise at the bottom of the 1.1
$\micron$ H$_2$O and CH$_4$ absorption trough.  CIA H$_2$ is
responsible for the slope on the red side of the K-band peak.
These features are characteristic of mid- and late-type T dwarfs
\citep{me02,geb02}, and confirm 2MASS 1503+2525 as a bona-fide T
dwarf.

\section{The Spectral Type and Spectrophotometric Distance of 2MASS 1503+2525}

We derived the spectral type of 2MASS 1503+2525 using the
independent classification schemes of \citet{me02} and
\citet{geb02}.  Spectral indices for both schemes sample the major
H$_2$O and CH$_4$ bands in T dwarf spectra, while the scheme of
Burgasser et al.\ also includes color indices between the various
spectral peaks. Measured ratio values and individual subtype
determinations are listed in Table 3, following the procedures
outlined by these two classification schemes.  The substantial
scatter amongst the Geballe et al.\ indices is caused by sampling
low signal-to-noise regions in the Gemini Camera spectrum;
similarly, the 2.2$\micron$ CH$_4$, K/J
2.11$\micron$/2.07$\micron$ indices in the Burgasser et al.\
scheme are affected by the poor signal-to-noise in the K-band
spectrum.  Nonetheless, both yield a common spectral type of T5.5.

We confirm this spectral type in Figure 4 by comparing the J- and
H-band spectra of 2MASS 1503+2525 to UKIRT CGS4 \citep{mnt90}
spectral data of the T5 2MASS 0559$-$1404 \citep{me00c,geb02} and
the T6 SDSS 1624+0029 \citep{str99}. Adequate comparison can be
made between these data, as their spectral resolutions are similar
(${\lambda}/{\Delta}{\lambda}$ $\sim$ 400--500) and all have been
corrected for telluric absorption and flux calibrated. 2MASS
1503+2525 has stronger H$_2$O and CH$_4$ absorption bands than
2MASS 0559$-$1404, but weaker bands than SDSS 1624+0029,
consistent with a median type of T5.5.

As noted above, the colors of 2MASS 1503+2525 are quite blue,
for its assigned spectral type of T5.5,
suggestive of either a later spectral type or suppression of
K-band flux, as is observed in the possibly high surface gravity
or metal-poor T dwarf 2MASS 0937+2931 \citep{me02}.  However,
within the photometric uncertainties these colors are still
consistent with other mid-type T dwarfs such as the T5 2MASS
0755+2212 (J--H = 0.06$\pm$0.17, H--K$_s$ = $-$0.08$\pm$0.26) and
the T5.5 2MASS 1534$-$2952AB (J--H = 0.03$\pm$0.13, H--K$_s$ =
0.03$\pm$0.16; Burgasser et al.\ 2002, 2003).  Higher accuracy
photometry of a sample of mid-type T dwarfs is required to
determine if the colors of 2MASS 1503+2525 are unusually blue for
its spectral type.

Both of the comparison objects in Figure 4 have known distances
from trigonometric parallax measurements \citep{dah02}. Therefore,
we can estimate the spectrophotometric distance of 2MASS 1503+2525
from the absolute 2MASS magnitudes of these objects and data from
Table 3.  If we assume this object has a spectral
type and hence luminosity similar to 2MASS 0559$-$1404, it lies
between 10.7--12.0 pc from the Sun.  This estimate may be inflated
if 2MASS 0559$-$1404 is a multiple system, as suggested by
\citet{me01}\footnote{This object was not resolved in HST
observations, however, requiring the hypothetical bright companion
to have a projected separation less than 0.5 AU \citep{me02c}.}.
On the other hand, if 2MASS 1503+2525 is a later-type, less
luminous T dwarf similar to SDSS 1624+0029, it lies only 4.9--5.4
pc from the Sun. The mean spectrophotometric distance between
these extremes is 8$\pm$3 pc; hence, if it is single, 2MASS
1503+2525 is likely within the 10 pc RECONS distance, and is
potentially a member of the 8 pc sample of \citet{rei97}.
Trigonometric parallax observations are currently underway for
2MASS 1503+2525 through the USNO Parallax Program (C.\ Dahn,
priv.\ comm.).

Given its brightness, it is possible that 2MASS 1503+2525 could be
a binary brown dwarf, as roughly 20\% of late-M, L, and T dwarfs
identified in magnitude-limited searches are found to have binary
companions with separations $a \lesssim$ 10 AU
\citep{koe99,rei01,clo02,me02c}. The presence of a companion would
increase our distance estimate by up to a factor of 1.4. Gemini
images rule out a bright companion beyond 1$\farcs$2, or 6--13 AU
assuming on a distance of 5--11 pc.  Because the apparent
separation limit of brown dwarf binaries is similar to this
observational limit \citep{me02c}, we cannot rule out the
multiplicity of 2MASS 1503+2525 with any confidence;
higher-resolution imaging (AO or space-based) or radial velocity
monitoring are required.

\section{Discussion}

The possibility that 2MASS 1503+2525 lies within 10 pc of the Sun
raises the question of how many late-type dwarfs are currently
known in this volume.  Table 4 lists all L and T dwarfs identified
to date which are or may
potentially be within the RECONS 10 pc horizon, based on
trigonometric \citep{dah02,pry97} or spectrophotometric parallax\footnote{We
computed spectrophotometric
parallaxes for all L dwarfs
in the compilations of \citet{del97,del99}; \citet{kir97};
\citet{kir99,kir00}; \citet{mrt99};
\citet{fan00}; \citet{giz00}; \citet{leg00};
\citet{geb02}; \citet{giz02}; \citet{haw02}; \citet{lod02};
\citet{sch02}; \citet{cru03}; Tinney et al.\ (in preparation);
and Wilson et al.\ (in preparation);
and all T dwarfs from \citet{str99}; \citet{me99,me00c,me02};
\citet{leg00}; \citet{tsv00}; and \citet{geb02}. See
\url{http://spider.ipac.caltech.edu/staff/davy/ARCHIVE/index.html}
for a current list of all known L and T dwarfs.},
respectively. We have used published spectrophotometric parallaxes
for the L dwarfs \citep{cru03,kir00} or computed them using the
spectral type/absolute magnitude relations of \citet{dah02},
2MASS photometry,
and spectral classifications by \citet{kir99}. We
derived distances for the T dwarfs by interpolating over
spectral type (using the Burgasser et al.\ scheme) between the
2MASS J, H, and K$_s$ absolute magnitudes of the L8 Gliese 584C
\citep{kir00,pry97}, T2 SDSS 1254$-$0122 \citep{leg00,dah02}, T5
2MASS 0559$-$1404 \citep{me00c,dah02}, T6 SDSS 1624+0029
\citep{str99,dah02}, T6.5 Gliese 229B \citep[UKIRT JHK
photometry]{leg99,pry97}, and T8 Gliese 570D \citep{me00a,pry97}.
The average of spectrophotometric distances derived from
2MASS JHK$_s$ photometry are reported.

While there are currently 313 known stars spectral type M or
earlier in 227 systems within 10 pc of the Sun (T.\ Henry, priv.\
comm.), only 5 L and 10 T dwarfs meet the same criteria, if we
assume that the spectrophotometric distances are accurate.  In
contrast, mass function simulations by \citet{me01}, assuming
conservatively $\Psi$(M) $\equiv$ dN/dM $\propto$ M$^{-0.5}$ for
0.01 M$_{\sun} \leq$ M $\leq 0.1$
M$_{\sun}$\footnote{\citet{cha02} find $\Psi$(M) $\propto$
M$^{-1}$, and \citet{rei99} find $\Psi$(M) $\propto$ M$^{-1.3}$,
over the same mass range.}, predict 12 L dwarfs (1300 $\lesssim$
T$_{eff}$ $\lesssim$ 2200 K) and 21 T dwarfs (800 $\lesssim$
T$_{eff}$ $\lesssim$ 1300 K) in the same spatial volume, more than
twice the number so far identified. The dearth of observed L and T
dwarfs worsens dramatically if the substellar mass function is
steeper.  The primary reason for this discrepancy is quite simple
- searches for these objects using large-scale surveys have not
yet examined the entire sky.  SDSS and DENIS programs have only
searched a fraction of their final survey area, which ultimately
will cover one-quarter and one-half of the sky, respectively.
Searches for L and T dwarfs in the all-sky 2MASS survey have only
examined roughly one-half of the sky.  Hence, the expanded areal
coverage and color criteria of our current search will likely
identify quite a few of the missing nearby T dwarfs, particularly
those with spectral types later than T4.  However, some of the
latest-type T dwarfs within 10 pc, and objects significantly less
luminous than Gliese 570D, may be too faint to be detected by
2MASS, SDSS, or DENIS, and will remain to be uncovered in future
surveys.

2MASS 1503+2525 is a key T dwarf discovery not just because of its
potential proximity to the Sun, but also because it is
sufficiently bright to permit both higher-resolution spectroscopic
study and investigations outside the 1--2.5 $\micron$ spectral
energy window. In particular, this object is a prime SIRTF target
for mid-infrared observations, a spectral region containing
diagnostics of temperature (e.g., NH$_3$ at 10.5 $\micron$) and
atmospheric composition (e.g., 6--9 $\micron$ silicate bands) that
is largely inaccessible due to the intrinsic faintness of these
cool brown dwarfs.

\acknowledgments

We thank our referee Sandy Leggett for her useful criticisms of
our manuscript and for providing the UKIRT spectral data of 2MASS
0559-1404 and SDSS 1624+0029 used in our analysis.  We acknowledge
useful discussions on the nearby star sample with Kelle Cruz and Todd
Henry. We also thank the Gemini Camera Instrument Specialist
Elinor Gates and Telescope Operator Andy Tullis at Lick for
assistance in the observations, and the UCO TAC for its allocation
of time for this project. Adam J.\ B.\ acknowledges support by NASA
through Hubble Fellowship grant HST-HF-01137.01 awarded by the
Space Telescope Science Institute, which is operated by the
Association of Universities for Research in Astronomy, Inc., for
NASA, under contract NAS 5-26555. This research has made use of
the SIMBAD database, operated at CDS, Strasbourg, France. POSS-I
and POSS-II images were obtained from the Digitized Sky Survey
image server maintained by the Canadian Astronomy Data Centre,
which is operated by the Herzberg Institute of Astrophysics,
National Research Council of Canada. This research has made use of
the NASA/IPAC Infrared Science Archive, which is operated by the
Jet Propulsion Laboratory, California Institute of Technology,
under contract with the National Aeronautics and Space
Administration. This publication makes use of data from the Two
Micron All Sky Survey, which is a joint project of the University
of Massachusetts and the Infrared Processing and Analysis Center,
funded by the National Aeronautics and Space Administration and
the National Science Foundation. Adam J.\ B.\ dedicates this
publication to the memory of Dr.\ Charles Kincaid Witham.

\clearpage


\begin{figure}
\plotone{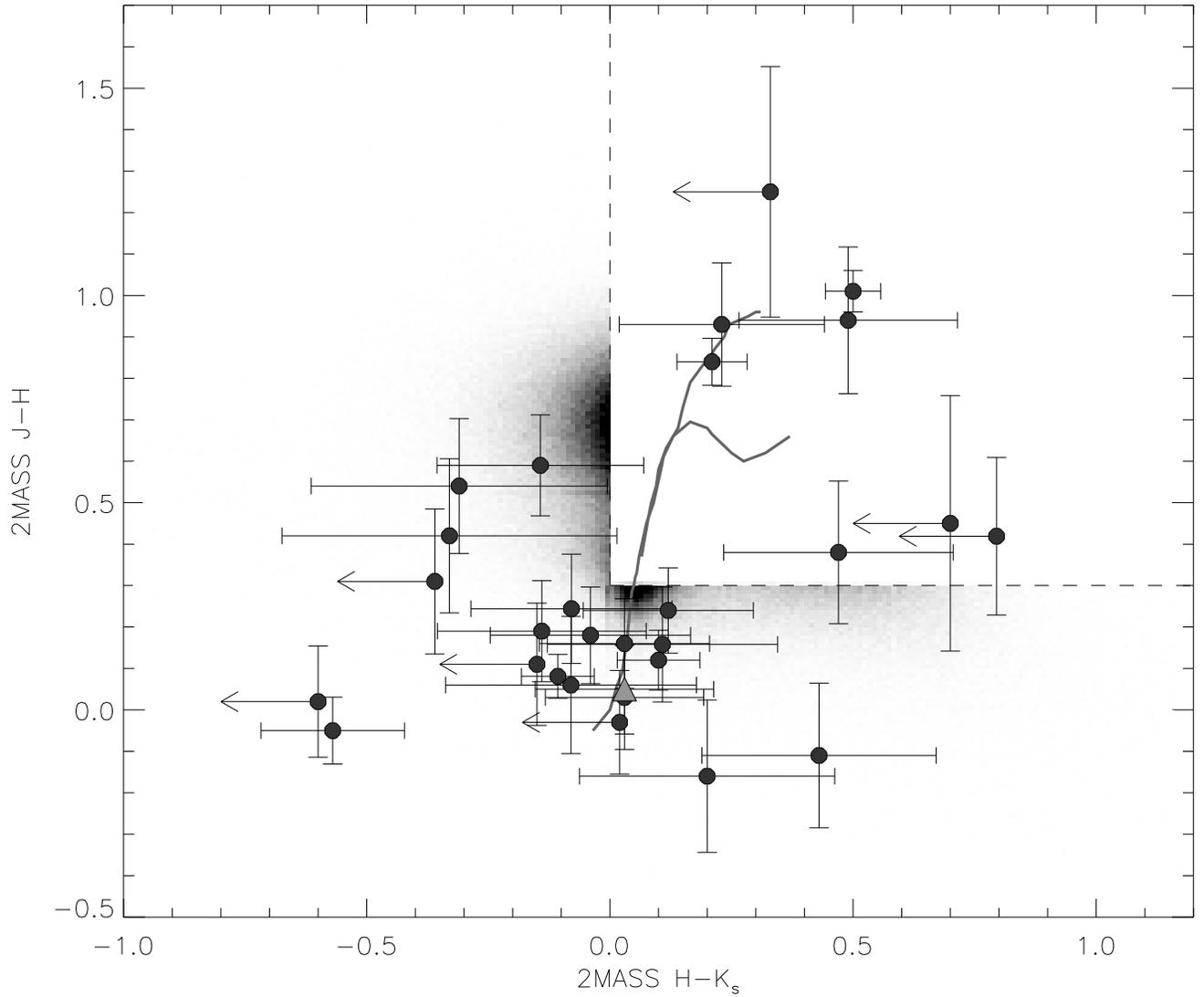} \caption{2MASS J--H, H--K$_s$ Hess
diagram of our T dwarf candidates, with the darkest regions
indicating the highest source densities. Color constraints are
indicated by dashed lines. The colors of 28 T dwarfs identified by
2MASS and SDSS are plotted as circles with error bars; arrows
indicate upper limits. The colors of 2MASS 1503+2525 are
distinguished by the large grey triangle. The \citet{bes88} giant
and dwarf tracks are delineated by the grey lines extending
redward of our sample color limits.\label{fig1}}
\end{figure}

\begin{figure}
\plotone{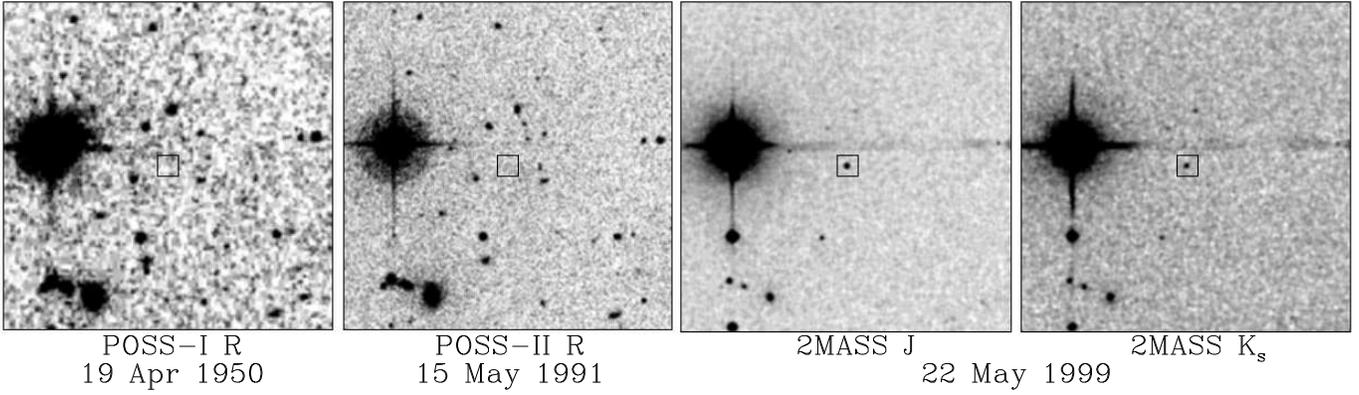} \caption{The field of 2MASS
1503+2525, from left to right: POSS-I (R-band), POSS-II (R-band),
and 2MASS (J- and K$_s$-bands).  Images are scaled to the same
spatial resolution, 5$\arcmin$ on a side, with North up and East
to the left.  A 10$\arcsec$ box is centered on the position of
2MASS 1503+2525 in all images.  The bright star 18$\arcsec$ North
and 103$\arcsec$ East of the T dwarf is the M5 variable giant EX
Boo \citep{gug97}. Note that two bright sources below EX Boo in
the 2MASS images are residual persistence sources caused by that
bright star (Cutri et al.\ 2002). \label{fig2}}
\end{figure}

\begin{figure}
\plotone{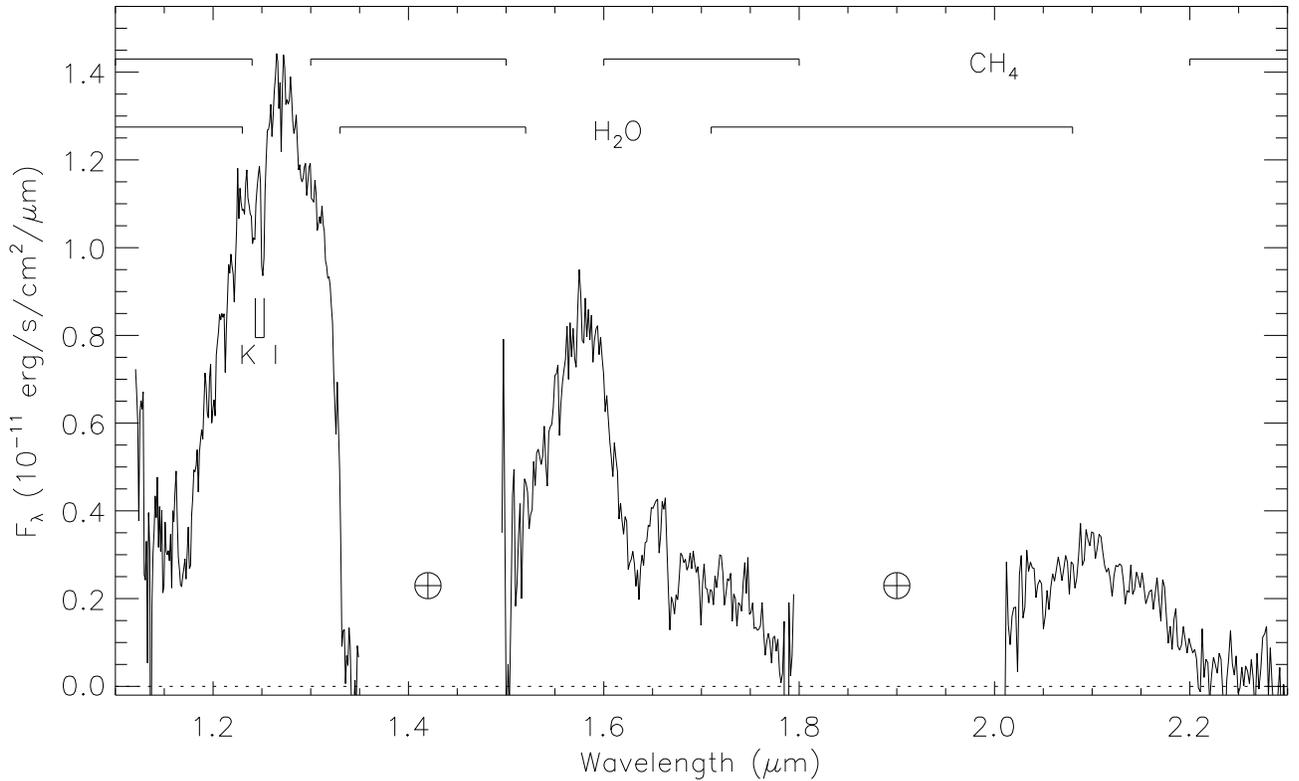} \caption{1.1--2.3 $\micron$ spectrum
of 2MASS 1503+2525 obtained with the Lick Gemini Camera.  CH$_4$,
H$_2$O, CIA H$_2$, and K I absorption features are labelled.
Regions of significant telluric absorption are indicated by the
$\oplus$ symbols. \label{fig3}}
\end{figure}

\begin{figure}
\plotone{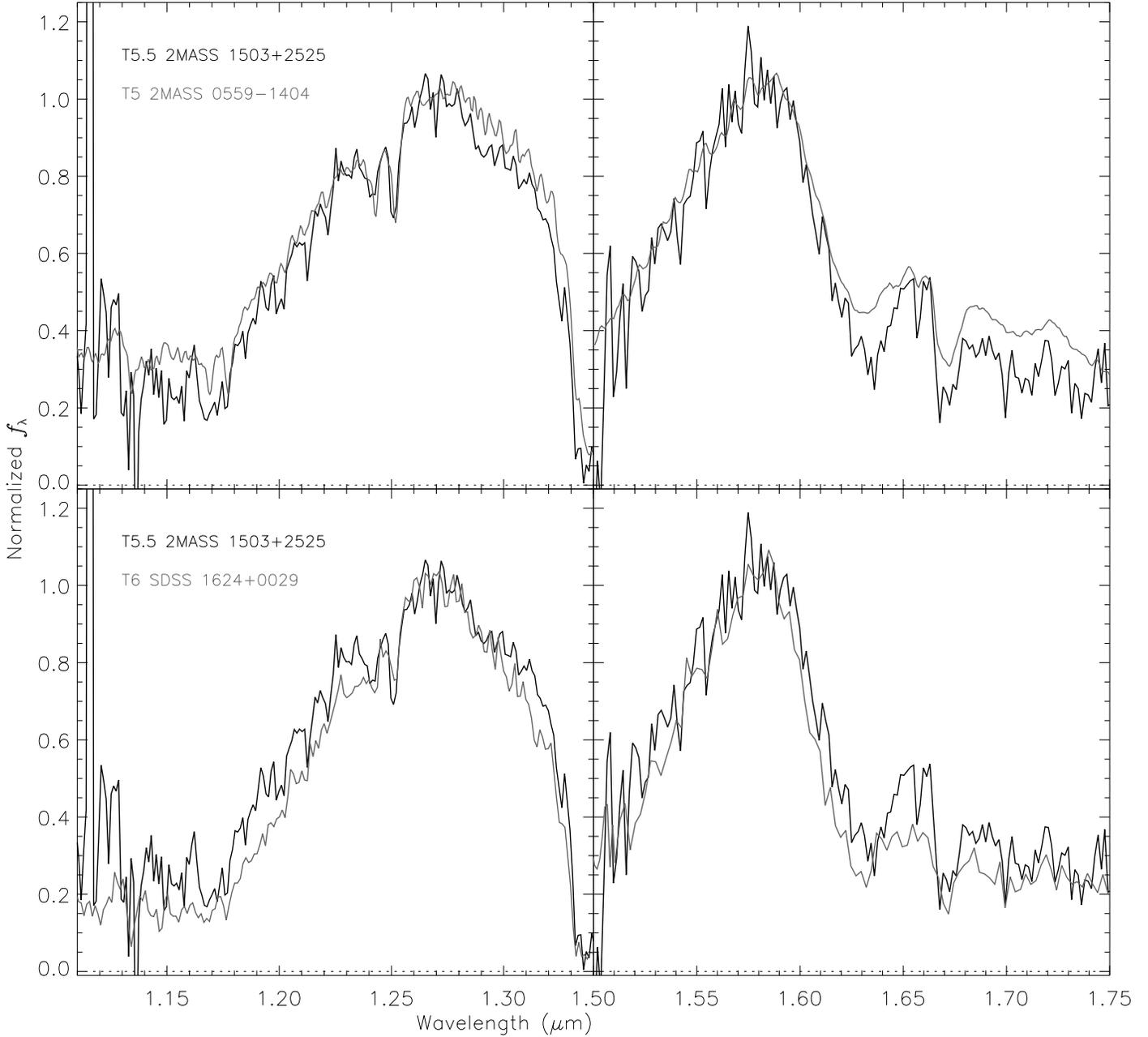} \caption{Comparison of the J- (left)
and H- (right) band spectra of (black lines) 2MASS 1503+2525 to
(grey lines) 2MASS 0559-1404 (top) and SDSS 1624+0029 (bottom).
Comparison spectral data, measured with the UKIRT CGS4 instrument,
are from \citet{geb02} and \citet{str99}, respectively.  All
spectra are normalized at their flux peaks in each band.
\label{fig4}}
\end{figure}

\clearpage

\begin{deluxetable}{llcccc}
\tabletypesize{\scriptsize} \tablecaption{Known T Dwarfs in the
2MASS Wide-Field Sample. \label{tbl-0}} \tablewidth{0pt}
\tablehead{ \colhead{Object\tablenotemark{a}} &
\colhead{SpT\tablenotemark{b}} & \colhead{2MASS J} &
\colhead{2MASS H} & \colhead{2MASS K$_s$} &
\colhead{Ref}   \\
\colhead{(1)} & \colhead{(2)} & \colhead{(3)} & \colhead{(4)} &
\colhead{(5)} & \colhead{(6)} } \startdata
2MASS J02431371$-$2453298 & T6  & 15.38$\pm$0.06 & 15.14$\pm$0.12 & 15.22$\pm$0.17 & 1 \\
2MASS J04151954$-$0935066 & T8  & 15.70$\pm$0.07 & 15.54$\pm$0.12 & 15.43$\pm$0.20 & 1 \\
2MASS J05591914$-$1404488 & T5  & 13.80$\pm$0.04 & 13.68$\pm$0.06 & 13.58$\pm$0.06 & 2 \\
2MASS J07271824+1710012 & T7  & 15.60$\pm$0.07 & 15.76$\pm$0.17 & 15.56$\pm$0.20 & 1 \\
2MASS J07554795+2212169 & T5:  & 15.73$\pm$0.07 & 15.67$\pm$0.15 & 15.75$\pm$0.21 & 1 \\
SDSSp J092615.38+584720.9 & T4.5  & 15.90$\pm$0.07 & 15.31$\pm$0.10 & 15.45$\pm$0.19 & 3 \\
2MASS J09373487+2931409 & T6p  & 14.65$\pm$0.04 & 14.70$\pm$0.07 & 15.27$\pm$0.13 & 1 \\
2MASS J10475385+2124234 & T6.5  & 15.82$\pm$0.06 & 15.80$\pm$0.12 & $>$ 16.4 & 4 \\
2MASS J12171110$-$0311131 & T7.5  & 15.86$\pm$0.07 & 15.75$\pm$0.13 & $>$ 15.9 & 4 \\
2MASS J12255432$-$2739466AB & T6  & 15.26$\pm$0.06 & 15.10$\pm$0.09 & 15.07$\pm$0.15 & 4,5 \\
2MASS J12373919+6526148\tablenotemark{c} & T6.5  & 15.81$\pm$0.07 & 16.28$\pm$0.24 & $>$ 15.9 & 4 \\
SDSSp J134646.45$-$003150.4 & T6  & 16.00$\pm$0.11 & 15.46$\pm$0.12 & 15.77$\pm$0.28 & 6 \\
Gliese 570D & T8  & 15.32$\pm$0.06 & 15.27$\pm$0.09 & 15.24$\pm$0.16 & 7 \\
2MASS J15031961+2525196 & T5.5  & 13.94$\pm$0.03 & 13.86$\pm$0.04 & 13.96$\pm$0.06 & 8 \\
2MASS J15344984$-$2952274AB & T5.5  & 14.90$\pm$0.06 & 14.87$\pm$0.11 & 14.84$\pm$0.12 & 1,5 \\
2MASS J15462718$-$3325111 & T5.5  & 15.63$\pm$0.06 & 15.45$\pm$0.10 & 15.49$\pm$0.18 & 1 \\
2MASS J15530228+1532369 & T7  & 15.83$\pm$0.07 & 15.94$\pm$0.16 & 15.51$\pm$0.18 & 1 \\
SDSSp J162414.37+002915.6 & T6  & 15.49$\pm$0.06 & 15.52$\pm$0.11 & $>$ 15.5 & 9 \\
2MASS J22541892+3123498 & T5  & 15.26$\pm$0.05 & 15.02$\pm$0.09 & 14.90$\pm$0.15 & 1 \\
2MASS J23391025+1352284\tablenotemark{c} & T5.5  & 15.88$\pm$0.08 & 16.13$\pm$0.18 & $>$ 15.5 & 1 \\
2MASS J23565477$-$1553111 & T6  & 15.82$\pm$0.07 & 15.63$\pm$0.10 & 15.77$\pm$0.19 & 1 \\
\enddata
\tablenotetext{a}{Designations (for 2MASS discoveries) and
photometry from the 2MASS PSC.} \tablenotetext{b}{Spectral types
from \citet{me02} except for SDSS 0926+5847 \citep{geb02}.}
\tablenotetext{c}{Photometry from another epoch satisfy our color
and magnitude constraints; see Table 2.} \tablerefs{(1)
\citet{me02}; (2) \citet{me00c}; (3) \citet{geb02}; (4)
\citet{me99}; (5) \citet{me02c}; (6) \citet{tsv00}; (7)
\citet{me00a}; (8) This paper; (9) \citet{str99}.}
\end{deluxetable}

\begin{deluxetable}{lccccccccc}
\tabletypesize{\scriptsize} \tablecaption{Multiple T Dwarf
Detections in the 2MASS WPSD. \label{tbl-01b}} \tablewidth{0pt}
\tablehead{ \colhead{Object}  & \colhead{SpT} &
\colhead{Epoch\tablenotemark{a}} & \colhead{$\alpha$ (J2000)} &
\colhead{$\delta$ (J2000)} & \colhead{2MASS J} & \colhead{2MASS H}
& \colhead{2MASS K$_s$} &
\colhead{$\mu$ (yr$^{-1}$)\tablenotemark{b}} & \colhead{$\theta$ ($\degr$)\tablenotemark{b}}   \\
\colhead{(1)} & \colhead{(2)} & \colhead{(3)} & \colhead{(4)} &
\colhead{(5)} & \colhead{(6)} & \colhead{(7)} & \colhead{(8)} &
\colhead{(9)} & \colhead{(10)}} \startdata
SDSS 0926+5847 & T4.5 & 990422 & 09$^h$26$^m$15$\fs$39 & +58$\degr$47$\arcmin$21$\farcs$19 & 15.60$\pm$0.08 & 15.62$\pm$0.17 & 15.41$\pm$0.24 &  0$\farcs$3$\pm$0$\farcs$3 & 220$\pm$40  \\
 & & 990422 & 09$^h$26$^m$15$\fs$43 & +58$\degr$47$\arcmin$21$\farcs$34 & 15.74$\pm$0.09 & 15.64$\pm$0.18 & $>$ 16.0 &  &  \\
 & & 000320 & 09$^h$26$^m$15$\fs$40 & +58$\degr$47$\arcmin$20$\farcs$90 & 15.75$\pm$0.07 & 15.33$\pm$0.10 & 15.46$\pm$0.17 & & \\
 & & 000320\tablenotemark{c} & 09$^h$26$^m$15$\fs$38 & +58$\degr$47$\arcmin$21$\farcs$23 & 15.90$\pm$0.07 & 15.31$\pm$0.10 & 15.45$\pm$0.19 & & \\
2MASS 0937+2931 & T6p & 981124 & 09$^h$37$^m$34$\fs$81 & +29$\degr$31$\arcmin$42$\farcs$66 & 14.66$\pm$0.04 & 14.68$\pm$0.08 & 15.52$\pm$0.24 & 1$\farcs$4$\pm$0.2 & 150$\pm$4 \\
 & & 000402\tablenotemark{c} & 09$^h$37$^m$34$\fs$88 & +29$\degr$31$\arcmin$40$\farcs$99 & 14.65$\pm$0.04 & 14.70$\pm$0.07 & 15.27$\pm$0.13 & & \\
2MASS 1237+6526 & T6.5 & 990313 & 12$^h$37$^m$39$\fs$18 & +65$\degr$26$\arcmin$14$\farcs$59 & 15.81$\pm$0.07 & 16.28$\pm$0.24 & $>$ 15.9 & ---\tablenotemark{d} & ---\tablenotemark{d} \\
 & & 990313\tablenotemark{c} & 12$^h$37$^m$39$\fs$20 & +65$\degr$26$\arcmin$14$\farcs$81 & 16.05$\pm$0.09 & 15.74$\pm$0.15 & $>$ 16.1 & & \\
SDSS 1346$-$0031 & T6 & 990122 & 13$^h$46$^m$46$\fs$43 & $-$00$\degr$31$\arcmin$50$\farcs$43 & 15.89$\pm$0.08 & 16.02$\pm$0.21 & $>$ 15.7 & 0$\farcs$5$\pm$0$\farcs$3 & 275$\pm$15 \\
 & & 000604 & 13$^h$46$^m$46$\fs$43 & $-$00$\degr$31$\arcmin$50$\farcs$33 & 16.00$\pm$0.12 & 15.91$\pm$0.26 & 15.44$\pm$0.23 & & \\
 & & 010203\tablenotemark{c} & 13$^h$46$^m$46$\fs$34 & $-$00$\degr$31$\arcmin$50$\farcs$13 & 16.00$\pm$0.11 & 15.46$\pm$0.12 & 15.77$\pm$0.28 & & \\
 & & 010208 & 13$^h$46$^m$46$\fs$38 & $-$00$\degr$31$\arcmin$50$\farcs$55 & 16.12$\pm$0.12 & 15.88$\pm$0.23 & $>$ 15.6 & & \\
Gliese 570D & T8 & 980516\tablenotemark{c} & 14$^h$57$^m$14$\fs$96 & $-$21$\degr$21$\arcmin$47$\farcs$75 & 15.32$\pm$0.06 & 15.27$\pm$0.09 & 15.24$\pm$0.16 & 1$\farcs$9$\pm$0$\farcs$2 & 149$\pm$3 \\
 & & 990729 & 14$^h$57$^m$15$\fs$04 & $-$21$\degr$21$\arcmin$49$\farcs$72 & 15.08$\pm$0.06 & 15.35$\pm$0.14 & 15.30$\pm$0.21 & & \\
2MASS 1534$-$2952AB & T5.5 & 980702 & 15$^h$34$^m$49$\fs$84 & $-$29$\degr$52$\arcmin$27$\farcs$40 & 14.92$\pm$0.06 & 14.81$\pm$0.09 & 14.98$\pm$0.13 & 0$\farcs$31$\pm$0$\farcs$12 & 137$\pm$11 \\
 & & 980708\tablenotemark{c}  & 15$^h$34$^m$49$\fs$84 & $-$29$\degr$52$\arcmin$27$\farcs$42 & 14.90$\pm$0.06 & 14.87$\pm$0.11 & 14.84$\pm$0.12 & & \\
 & & 000430 & 15$^h$34$^m$49$\fs$87 & $-$29$\degr$52$\arcmin$27$\farcs$82 & 14.97$\pm$0.07 & 14.72$\pm$0.07 & 14.85$\pm$0.12 & & \\
2MASS 2339+1352 & T5.5 & 980929 & 23$^h$39$^m$10$\fs$21 & +13$\degr$52$\arcmin$30$\farcs$20 & 15.88$\pm$0.08 & 16.13$\pm$0.18 & $>$ 15.5 & 0$\farcs$83$\pm$0$\farcs$11 & 159$\pm$11 \\
 & & 001029 & 23$^h$39$^m$10$\fs$23 & +13$\degr$52$\arcmin$28$\farcs$56 & 16.49$\pm$0.15 & 15.83$\pm$0.24 & $>$ 15.6 & & \\
 & & 001110 & 23$^h$39$^m$10$\fs$28 & +13$\degr$52$\arcmin$28$\farcs$64 & 16.18$\pm$0.11 & 15.70$\pm$0.17 & $>$ 15.8 & & \\
 & & 001129\tablenotemark{c} & 23$^h$39$^m$10$\fs$25 & +13$\degr$52$\arcmin$28$\farcs$48 & 16.24$\pm$0.11 & 15.82$\pm$0.15 & 16.15$\pm$0.31 & & \\
\enddata
\tablenotetext{a}{2MASS observation date in YYMMDD (UT).}
\tablenotetext{b}{Proper motions derived from separate linear fits
to Right Ascension and declination motion; uncertainties include
the fit uncertainties and 2MASS 0$\farcs$3 positional uncertainty.
Values have not been corrected for Solar motion.}
\tablenotetext{c}{Epoch used in 2MASS PSC.}
\tablenotetext{d}{Epoch difference too small to compute proper
motion.}
\end{deluxetable}

\begin{deluxetable}{lcc}
\tabletypesize{\scriptsize} \tablecaption{T Dwarf Spectral
Classification Indices for 2MASS 1503+2525} \tablewidth{0pt}
\tablehead{ \colhead{Index} & \colhead{Burgasser et
al.\tablenotemark{a}} & \colhead{Geballe et al.\tablenotemark{b}}
\\ \colhead{(1)} & \colhead{(2)} & \colhead{(3)} } \startdata
1.1$\micron$ H$_2$O & 0.274 (5--6) & 4.62 (6) \\
1.3$\micron$ CH$_4$ & 0.784 (5) & --- \\
1.5$\micron$ H$_2$O & 0.432 (5) & 3.28 (3) \\
1.6$\micron$ CH$_4$ & 0.364 (5--6) & 2.60 (6) \\
2.2$\micron$ CH$_4$ & 0.094 (7) & 10.8 (7) \\
H/J & 0.383 (5) & --- \\
K/J & 0.131 (8) & --- \\
2.11$\micron$/2.07$\micron$ & 1.12 (--) & --- \\
\hline
Final Spectral Type & T5.5 & T5.5 \\
\enddata
\tablenotetext{a}{Index subtypes determined by closest match to
spectral standard values; final subtype is the average of the
index subtypes after rejecting high and low values.}
\tablenotetext{b}{Index subtypes determined from predetermined
ranges; final subtype is the average of the index subtypes.}
\end{deluxetable}

\begin{deluxetable}{llccccc}
\tabletypesize{\scriptsize} \tablecaption{Known L and T Dwarfs
Within 10 pc of the Sun. \label{tbl-2}} \tablewidth{0pt}
\tablehead{ \colhead{Object} & \colhead{SpT\tablenotemark{a}}   &
\colhead{2MASS J}   & \colhead{2MASS H} & \colhead{2MASS K$_s$} &
\colhead{d (pc)\tablenotemark{b}} & \colhead{Ref.} } \startdata
2MASS 0036+1821 & L3.5  & 12.44$\pm$0.04 & 11.58$\pm$0.03 & 11.03$\pm$0.03 & 8.76$\pm$0.06 & 1 \\
2MASS 0835$-$0819 & L4.5 & 13.17$\pm$0.04 & 11.94$\pm$0.04 & 11.14$\pm$0.04 & {\em 9.1} & 2 \\
2MASS 1507$-$1627 & L5  & 12.82$\pm$0.03 & 11.90$\pm$0.03 & 11.30$\pm$0.03 & 7.33$\pm$0.03 & 1 \\
GJ 1001B & L5  & 13.10$\pm$0.03 & 12.05$\pm$0.02 & 11.40$\pm$0.03 & 9.55$\pm$0.10 & 3 \\
DENIS 0255$-$4700 & L8  & 13.23$\pm$0.03 & 12.19$\pm$0.02 & 11.53$\pm$0.03 & {\em 5.0} & 4 \\
SDSS 0423$-$0414 & T0  & 14.47$\pm$0.04 & 13.46$\pm$0.06 & 12.93$\pm$0.04 & {\em 8.9} & 5 \\
2MASS 1503+2525 & T5.5  & 13.94$\pm$0.03 & 13.86$\pm$0.04 & 13.96$\pm$0.06 & {\em 7.8} & 5 \\
2MASS 0937+2931 & T6p  & 14.65$\pm$0.04 & 14.70$\pm$0.07 & 15.27$\pm$0.13 & {\em 8.8} & 5 \\
2MASS 0243$-$2453 & T6  & 15.38$\pm$0.06 & 15.14$\pm$0.12 & 15.22$\pm$0.17 & {\em 9.8} & 5 \\
Gliese 229B & T6.5  & 14.32$\pm$0.05\tablenotemark{c} & 14.35$\pm$0.05\tablenotemark{c} & 14.42$\pm$0.05\tablenotemark{c} & 5.77$\pm$0.04 & 3 \\
2MASS 0727+1710 & T7  & 15.60$\pm$0.07 & 15.76$\pm$0.17 & 15.56$\pm$0.20 & {\em 9.1} & 5 \\
2MASS 1553+1532 & T7  & 15.83$\pm$0.07 & 15.94$\pm$0.16 & 15.51$\pm$0.18 & {\em 9.6} & 5 \\
2MASS 1217$-$0311 & T7.5  & 15.86$\pm$0.07 & 15.75$\pm$0.13 & $>$ 15.9 & {\em 8.6} & 5 \\
Gliese 570D & T8  & 15.32$\pm$0.06 & 15.27$\pm$0.09 & 15.24$\pm$0.16 & 5.91$\pm$0.06 & 3 \\
2MASS 0415$-$0935 & T8  & 15.70$\pm$0.07 & 15.54$\pm$0.12 & 15.43$\pm$0.20 & {\em 6.7} & 5 \\
\enddata
\tablenotetext{a}{Spectral types from \citet{kir99,kir00} and
\citet{me02}, except for 2MASS 0835$-$0819 \citep{cru03} and SDSS
0423$-$0414 \citep{geb02}.} \tablenotetext{b}{Measured values are
listed with uncertainties; spectrophotometric distance estimates
are listed in italics.} \tablenotetext{c}{UKIRT photometry from
\citet{leg99}.} \tablerefs{(1) \citet{dah02}; (2) \citet{cru03};
(3) \citet{pry97}; (4) \citet{kir00}; (5) This paper.}
\end{deluxetable}

\end{document}